\documentclass[11pt, onecolumn]{IEEEtran}
\pdfoutput=1
\usepackage[T1]{fontenc}
\usepackage{amsmath}
\usepackage{cancel}
\usepackage{amssymb,stmaryrd}
\usepackage{graphicx}
\usepackage{psfrag, setspace}
\usepackage{latexsym}
\usepackage{dsfont}
\usepackage{cite}
\usepackage{extraipa}
\usepackage{esvect}
\usepackage{mathrsfs}
\usepackage{color}
\usepackage{srcltx}
\usepackage{relsize,mdframed,xfrac}
\usepackage{eufrak,feyn}
\usepackage{float, array, makecell}

\newtheorem{theorem}{Theorem}

\newtheorem{lemma}{Lemma}

\newtheorem{definition}{Definition}

\newcommand{\qed}{\nobreak \ifvmode \relax \else
      \ifdim\lastskip<1.5em \hskip-\lastskip
      \hskip0.5em plus0em minus0.5em \fi \nobreak
      \vrule height0.4em width0.5em depth0.1em\fi}
\DeclareMathAlphabet{\mathpzc}{OT1}{pzc}{m}{it}

\def\stackrel#1#2{\mathrel{\mathop{#2}\limits^{#1}}}

\newcommand{\set}[1]{{\mathcal{#1}}}

\newcommand{\mc}[1]{{\mathcal{#1}}}

\newcommand{\ve}{{\varepsilon}}
\DeclareMathOperator{\Exp}{{\mathbb{E}}} 
\newcommand{\U}{{\set{U}}} 
\newcommand{\X}{{\set{X}}} 
\newcommand{\Y}{{\set{Y}}} 
\newcommand{\supp}{\mathsf{supp}}

\renewcommand{\Pr}{\mathbb{P}}

\newcommand{\h}{\mathsf{h}}

\newcommand{\A}{{\set{A}}} 
\newcommand{\B}{{\set{B}}} 

\definecolor{lgray}{gray}{0.45}
\newcommand{\mkv}{\color{lgray}\,\feyn{fs}\,\color{black}} 

\allowdisplaybreaks[4]
\title{\huge{Lossy Compression with Near-uniform Encoder Outputs}}
\author{\IEEEauthorblockN{Badri N. Vellambi,   J\"{o}rg Kliewer} \thanks{This work is supported by NSF grants CCF-1440014, CCF-1439465, CCF-1320304, and CCF-1527074. 
}

\IEEEauthorblockA{New Jersey Institute of Technology, 
Newark, NJ 07102\\
Email: \{badri.vellambi, jkliewer\}@njit.edu\\}
\and 
\and
\IEEEauthorblockN{Matthieu R.~Bloch}

\IEEEauthorblockA{Georgia Institute of Technology,
Atlanta, GA 30332\\
Email: matthieu.bloch@ece.gatech.edu}}
\IEEEoverridecommandlockouts
\begin{document}
\maketitle

\begin{abstract}
It is well known that lossless compression of a discrete memoryless source with near-uniform encoder output is possible at a rate above its entropy if and only if the encoder is randomized. This work focuses on deriving conditions for near-uniform encoder output(s) in the Wyner-Ziv and the distributed lossy compression problems. We show that in the Wyner-Ziv problem, near-uniform encoder output and operation close to the WZ-rate limit is simultaneously possible, whereas in the distributed lossy compression problem, jointly near-uniform outputs is achievable in the interior of the distributed lossy compression rate region if the sources share non-trivial G\'{a}cs-K\"{o}rner common information.
\end{abstract}
\section{Introduction}
Owing to the source-channel separation theorem for point-to-point communication and the convenience separation offers, separate source and channel coding and the optimality of separation have been studied in several multi-user problems. Separation-based approaches, especially in multi-user settings, usually assume that the output of source encoders are near-uniform in its alphabet, where uniformity is measured using the variational distance metric. While the lack of near-uniform encoder output(s) does not necessarily cause separation-based approaches to fail, a characterization of when compression of sources can be achieved with near-uniform encoder output(s) simplifies the analysis of separation-based schemes, and is certainly valuable from a theoretical perspective.

Lossless compression with vanishing error probability and near-uniform encoder output was explored in~\cite{Han-IT2005, Hayashi-IT2008}. Hayashi showed that vanishing error probability and near-uniform encoder output cannot be simultaneously achieved~\cite{Hayashi-IT2008}. However, one can design lossless codes with near-uniform encoder output if the encoder and decoder share a random seed whose size is roughly the square root of the blocklength of the code~\cite{Chou-Bloch-ISIT13, Vellambi-Kliewer-Bloch-ISIT15}. In \cite{Vellambi-Kliewer-Bloch-ISIT15}, we have also shown using finite-length results of Kontoyiannis et al.~\cite{Kontoyiannis-PtwiseRedundancy} that lossy compression arbitrarily close to the rate-distortion limit is possible even with near-uniform encoder output. In this work, we analyze the rate points for the Wyner-Ziv (WZ) and distributed lossy compression problems at which compression with near-uniform encoder output(s) is possible. Specifically, we have proven the following results.\\
\noindent $\bullet$ {\underline{\textit{Wyner-Ziv Problem:}}} Lossy compression with near-uniform encoder output is possible at all rates above the WZ-rate limit.

\noindent $\bullet$ {\underline{\textit{Two-source Distributed Lossy Compression Problem:}}}  If the sources share non-trivial G\'{a}cs-K\"{o}rner common information, then lossy compression with jointly near-uniform encoder outputs is achievable at any rate pair in the interior of the distributed lossy compression rate region. The case where the sources share no G\'{a}cs-K\"{o}rner common information is open.

The proofs for both problems employ ideas from channel resolvability~\cite[p.~404]{TSH-Book} and the likelihood encoder~\cite{Cuff-LikelihoodEnc}. The result for the distributed lossy compression case is proven without needing a characterization of the underlying rate region. Instead, we exploit the existence of codes with near-uniform encoder outputs for a variant of the Slepian-Wolf problem with a non-standard decoding constraint. 

The remainder of the paper is organized thus. Section~\ref{sec-Notation} provides the notation, Section~\ref{sec-PD} defines the problems studied, Section~\ref{Sec-Res} details the main results of this work, and lastly, Section~\ref{Sec-ReqRes} presents the results used in the proofs of Section~\ref{Sec-Res}.
 
\section{Notation}\label{sec-Notation}
For $m,n\in\mathbb{N}$ with $m<n$, $\llbracket m, n\rrbracket \triangleq\{m, m+1,\ldots,n\}$. Uppercase letters (e.g., $X$, $Y$) denote random variables, lower cases denote their realizations (e.g., $x$, $y$), and the respective script versions (e.g., $\X$, $\Y$) denote their alphabets. In this work, all alphabets are assumed to be finite. Superscripts indicate the length of vectors, and subscripts indicate the component index. Given a finite set $\mathcal{S}$, $\mathsf{unif}(\mc S)$ denotes the uniform probability mass function (pmf) on $\mc S$.
Given a pmf $p_X$, $\supp(p_X)$ indicates the support of $p_X$, $p_X^{\otimes n}$ indicates the joint pmf of $n$ i.i.d random variables distributed according to $p_X$, and $T_\ve^n[p_X]$ denotes the set of $\ve$-strongly letter typical sequences of length $n$~\cite{Kramer-MUIF}. Given an event $E$, $\mathbb{P}(E)$ denotes the probability of its occurrence. Lastly, given two pmfs $p$ and $q$ over a set $\mathcal{X}$, the variational distance is denoted by 
\begin{align}
\mathbb{V}(p,q)\triangleq \sum_{x\in\mathcal{X}} |p(x)-q(x)|.
\end{align}

\section{Problem Definition} \label{sec-PD}
The lossy coding problems studied in this work impose a near-uniform encoder output constraint on the classical Wyner-Ziv and distributed lossy compression problems, and are formally defined here for the sake of completeness.
\begin{definition}\label{Def-WZ}
Let discrete memoryless sources $(X,Y)$ correlated according to pmf $\mathsf Q_{XY}$, a bounded distortion measure $d:\X\times \hat{\X} \rightarrow [0, d_{\max}]$, and $\Delta\in  [0, d_{\max}]$ be given. We say that Wyner-Ziv coding of the source $X$ with receiver side-information $Y$ at an average per-symbol distortion of $\Delta$ and is achievable with near-uniform encoder output at a rate $R\in\mathbb{R}^+$  if for every $\ve>0$, there exist an $n\in \mathbb{N}$, an encoding function $f_X: \X^n \rightarrow \llbracket 1, 2^{n(R+\ve)}\rrbracket$ and a reconstruction function $g_X: \llbracket 1, 2^{n(R+\ve)}\rrbracket \times \Y^n \rightarrow \hat \X^n$ at the receiver  such that 
\begin{align}
\mathbb{V}(Q_{f_X(X^n)}, \mathsf{unif}(\llbracket 1, 2^{n(R+\ve)}\rrbracket))&\leq \ve,\\
{\textstyle \sum_{i=1}^n} \Exp d(X_i, \hat X_i)  &\leq n(\Delta+\ve),
\end{align}
where $ Q_{f_X(X^n)}$ is the pmf of the encoder output $f_X(X^n)$ and $\hat X^n = g_X(f_X(X^n), Y^n)$ is the receiver reconstruction. 
\end{definition}
\begin{definition}\label{Def-DLC}
Let discrete memoryless sources $(X,Y)$ correlated according to pmf $\mathsf Q_{XY}$, bounded distortion measures $d_X:\X\times \hat{\X} \rightarrow [0, {d_x}_{\max}]$ and $d_Y:\Y\times \hat{\Y} \rightarrow [0, {d_y}_{\max}]$, and $\Delta_x\in [0, {d_x}_{\max}]$, $\Delta_y\in [0, {d_y}_{\max}]$ be given. We say that distributed lossy compression with jointly near-uniform encoder outputs and at average per-symbol distortions of $\Delta_x$ and $\Delta_y$ for sources $X$ and $Y$, respectively, is achievable at a rate pair $(R_x,R_y)\in {\mathbb{R}^+}^2$ if for every $\ve>0$, there exist an $n\in \mathbb{N}$, encoding functions $f_X: \X^n \rightarrow \llbracket 1, 2^{n(R_x+\ve)}\rrbracket$, $f_Y: \Y^n \rightarrow \llbracket 1, 2^{n(R_y+\ve)}\rrbracket$, and a reconstruction function $g_{XY}: \llbracket 1, 2^{n(R_x+\ve)}\rrbracket \times \llbracket 1, 2^{n(R_y+\ve)}\rrbracket\rightarrow \hat \X^n\times \hat \Y^n$ such that
\begin{align}
\mathbb{V}(Q_{f_X(X^n), f_Y(Y^n)}, Q_U)&\leq \ve,\\
{\textstyle\sum_{i=1}^n}  \Exp d(X_i, \hat X_i) &\leq n(\Delta_x+\ve),\\
{\textstyle\sum_{i=1}^n}  \Exp d(Y_i, \hat Y_i) &\leq n(\Delta_y+\ve),
\end{align}
where  $Q_U$ is the uniform pmf on $\llbracket 1, 2^{n(R_x+\ve)}\rrbracket\times \llbracket 1, 2^{n(R_y+\ve)}\rrbracket$, $Q_{f_X(X^n), f_Y(Y^n)}$ is the pmf of the outputs $f_X(X^n)$, $f_Y(Y^n))$ of the two encoders, and $(\hat X^n,\hat Y^n) = g_{XY}(f_X(X^n), f_Y(Y^n))$ are the receiver reconstructions. 
\end{definition}

\section{Main Results}\label{Sec-Res}
\subsection{Near-uniform Wyner-Ziv Coding}
\begin{theorem}\label{thm:WZ}
Near-uniform encoder output is achievable in the Wyner-Ziv problem at rates $R\geq \mathsf R_{WZ}(\Delta)$. 
\end{theorem}
\begin{IEEEproof}
The proof builds codes based on channel resolvability~\cite[p.~404]{TSH-Book} and the likelihood encoder~\cite{Cuff-LikelihoodEnc}, which allow us to track the distribution of the encoder output more readily than when using the covering lemma. We first pick a channel $Q_{W|X}$ such that: 
\begin{itemize}
\item the pmf $Q_{W | X} \mathsf Q_{XY}$ satisfies
\begin{align}
I(X;W|Y)= I(X; W)-I(Y; W)= \mathsf R_{WZ}(\Delta),
\end{align}
\item  $\Exp[d(X,f(W,Y)]\leq \Delta$ for some function $f$ of $(W,Y)$. 
\end{itemize}

Now, fix $\ve>0$, and let $R\triangleq I(X;W|Y) + 2\ve$, and $R' \triangleq I(W; Y)-\ve$. Let the codebook $\mc C$ comprising of $2^{n(R+R')}$ $\hat{X}$-codewords with each codeword selected i.i.d accroding to $Q_W^{\otimes n}$, where $Q_W$ is the marginal of $W$ derived from $Q_{W|X}\mathsf Q_X$.  We arrange the codewords of the random codebook $\mathcal{C}$ in a table of $2^{nR}$ rows and $2^{nR'}$ columns. 
Suppose that 
\begin{align}
(K,K')\sim  Q_{K,K'} &\triangleq \mathsf{unif}(\llbracket 1, 2^{nR}\rrbracket\times \llbracket 1, 2^{nR'}\rrbracket). \label{eqn-KK'defn}
\end{align}
 denotes the random pair of indices used to select the codewords from the codebook $\mathcal C$. Let $W^n(K,K')$ be selected and transmitted over the discrete memoryless channel (DMC) ${Q}_{X |W}$, and $\tilde{X}^n$ be the corresponding output, and let $\tilde{Y}^n$ be the output when $\tilde{X}^n$ is transmitted over the DMC $\mathsf Q_{Y|X}$. 
  \begin{figure}[h!]
   \centering
   \includegraphics[scale=0.75]{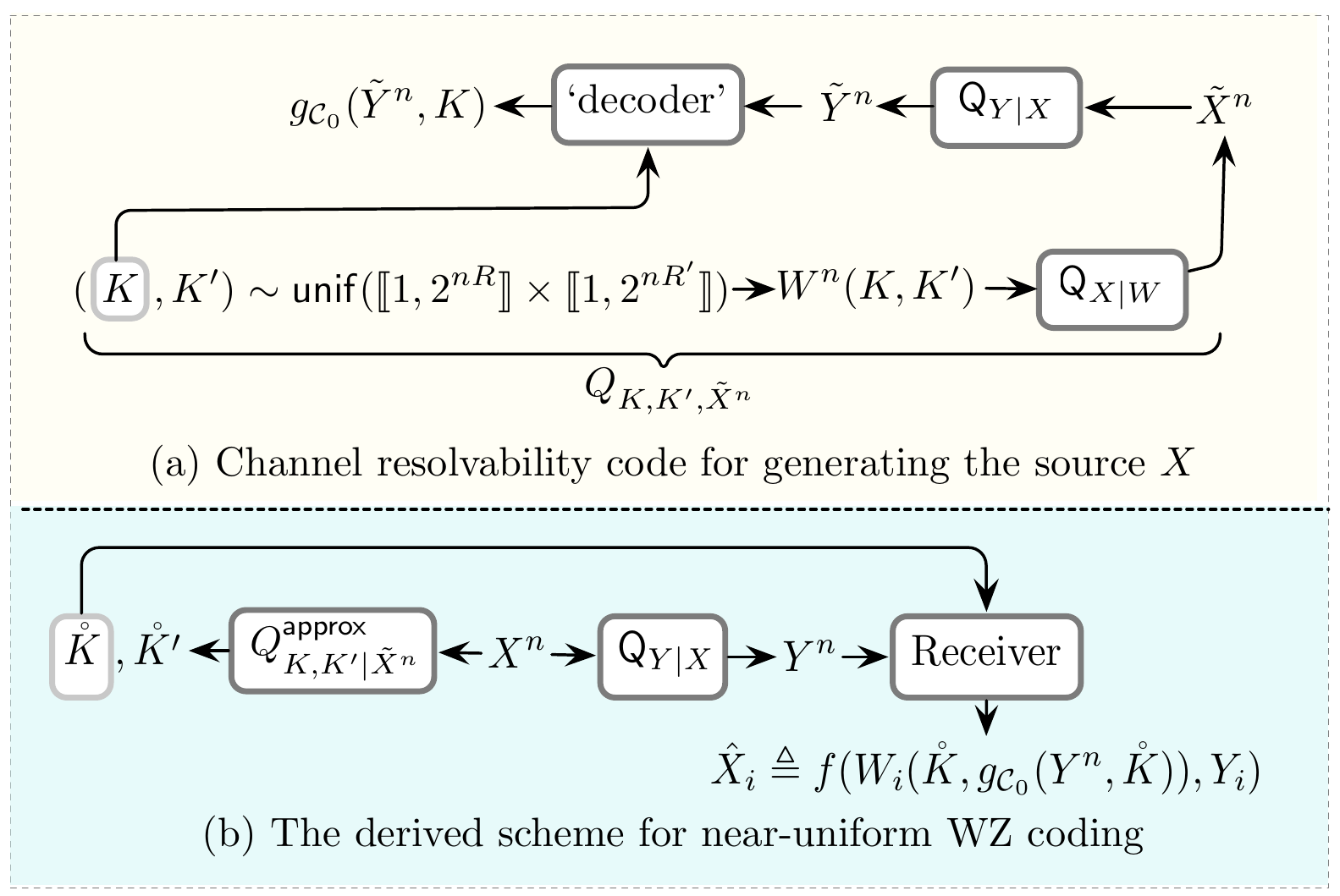}
   \caption{Source generation and the derived near-uniform WZ scheme.}\label{Fig:2-WZ}
 \end{figure}
 
 For this construction, the following hold:
 \begin{itemize}
 \item[1.] Since $R+R'> I (X; W)$, the channel resolvability theorem~\cite[Theorem 6.3.1]{TSH-Book} guarantees that 
\begin{align}
\Exp_\mathcal{C}[\mathbb{V} (Q_{\tilde X^n}, \mathsf Q_X^{\otimes n})] \stackrel{n\rightarrow \infty}{\longrightarrow} 0,
\end{align}
where the expectation is over all codebook realizations.
\item[2.] Since $R'<I(W; Y)$, there must exist a `decoding' function $g_\mathcal{C}: \Y^n \times \llbracket 1, 2^{nR}\rrbracket\rightarrow \llbracket 1, 2^{nR'}\rrbracket$ (depending on $\mathcal{C}$) such that 
\begin{align}
\Exp_\mathcal{C}\big[ \Pr[ K' \neq g_\mathcal{C}(\tilde{Y}^n, K) ] \big] \stackrel{n\rightarrow \infty}{\longrightarrow} 0.
\end{align}
\item[3.] Since $W^n(K,K')\sim Q_{W}^{\otimes n}$, and since $W^n(K,K')$ and $(\tilde{X}^n,\tilde Y^n)$  are related through the DMC $\mathsf Q_{Y|X}Q_{X|W}$, by the weak law of large numbers, we have
\begin{align}
\hspace{-1mm}\Pr\big[ (W^n(K,K'),\tilde{X}^n,\tilde Y^n) \hspace{-0.4mm}\notin\hspace{-0.4mm} T^n_\ve [ Q_{W|X}\mathsf Q_{XY}] \big] \stackrel{n\rightarrow \infty}{\longrightarrow} 0.
\end{align}
\end{itemize}
Now, for sufficiently large $n$, we can find a realization $\mathcal{C}_0=\{w_{\mc C_0}^n(j,k): j\in\llbracket 1, 2^{nR}\rrbracket, k\in \llbracket 1, 2^{nR'}\rrbracket\}$ of the codebook such that the sources $\tilde{X}^n$ and $\tilde Y^n$ generated by transmitting a codeword selected uniformly at random from $\mathcal{C}_0$ satisfy:
\begin{align}
\mathbb{V} (Q_{\tilde X^n}, \mathsf Q_X^{\otimes n}) &\leq \sfrac{\ve}{2} \label{eqn:WZ1}\\
 \Pr[ K' \neq g_{\mathcal{C}_0}(\tilde{Y}^n, K) ] &\leq \sfrac{\ve}{2}\label{eqn:WZ2}\\
 \Pr\big[ (W^n(K,K'),\tilde{X}^n,\tilde Y^n) \notin T^n_\ve [ Q_{W|X}\mathsf Q_{XY}] \big] &\leq \sfrac{\ve}{2}\label{eqn:WZ3}.
\end{align}
Let  $Q_{K,K' \tilde{X}^n}$ be the pmf induced by the codebook $\mathcal C_0$. Now, to derive a (randomized) WZ scheme from this channel resolvability code, we proceed as given in Fig.~\ref{Fig:2-WZ}. We first pick an approximation $Q_{K,K',\tilde X^n}^\textsf{\tiny{approx}}$ of $Q_{K,K',\tilde X^n}$ such that 
\begin{align}
\mathbb{V}( Q_{K,K',\tilde X^n}^\textsf{\tiny{approx}}, Q_{K,K', \tilde X^n})\leq \sfrac{\ve}{2}. \label{eqn:ApproxQ}
\end{align}
The need for an approximation will become clear later when we \emph{emulate} $Q_{K,K',\tilde X^n}$ using $\tilde X^n$ and a near-uniform random seed. Upon choosing $Q_{K,K',\tilde X^n}^\textsf{\tiny{approx}}$, we encode $X^n$ by generating $(\mathring K, \mathring K') \sim Q_{K,K'| \tilde X^n}^\textsf{\tiny{approx}}(\cdot,\cdot| X^n)$. We then transmit only $\mathring K$ to the receiver.  The joint pmf of $(\mathring K, \mathring K', X^n)$ is given by
\begin{align}
Q_{\mathring K, \mathring K', X^n}(\mathring k, \mathring k', x^n) \triangleq Q^\textsf{\tiny{approx}}_{ K,K'\mid \tilde X^n}(\mathring k, \mathring k'| x^n) \mathsf Q^{\otimes n} (x^n). \label{eqn:WZ4}
\end{align}
From \eqref{eqn:WZ1}, \eqref{eqn:ApproxQ} and \eqref{eqn:WZ4}, we are guaranteed that
\begin{align}
\mathbb{V}( Q_{\mathring K, \mathring K', X^n}, Q_{K,K',\tilde{X}^n}) 
&\leq\mathbb V(Q_{\mathring K, \mathring K', \tilde X^n}^{\textsf{\tiny{approx}}}, Q_{K,K', \tilde X^n})+ \mathbb{V} (Q_{\tilde X^n}, \mathsf Q_X^{\otimes n})\stackrel{}{\leq} \ve.
\end{align}
Further, since $Y^n$ and $\tilde Y^n$ are the outputs when $X^n$ and $\tilde{X}^n$, respectively, are fed into the DMC $\mathsf Q_{Y|X}$,  we are guaranteed to have
\begin{align}
\mathbb{V}( Q_{\mathring K, \mathring K', X^n, {Y}^n}, Q_{K,K',\tilde X^n, \tilde{Y}^n}) &\leq \ve.\label{eqn:WZ5}
\end{align}
Consequently, the following also hold
\begin{align}
\mathbb{V}( Q_{\mathring K, \mathring K', {Y}^n}, Q_{K,K',\tilde{Y}^n}) &\leq \ve\label{eqn:WZ5-1}\\
\mathbb{V}( Q_{\mathring K, \mathring K'}, Q_{K,K'}) &\leq \ve, \label{eqn:WZ5-2}
\end{align}
 From  \eqref{eqn-KK'defn} and \eqref{eqn:WZ5-2}, we see that  $\mathring{K}$ and $\mathring{K}'$ are jointly nearly uniform. Hence, $\mathring{K}$, which is the WZ encoder output, is also nearly uniform. Further, \eqref{eqn:WZ3} and \eqref{eqn:WZ5} jointly imply that
\begin{align}
 \Pr\big[ ( W^n(\mathring K,\mathring K'),X^n,Y^n) \notin T^n_\ve [ Q_{W|X}\mathsf Q_{XY}] \big]  &\stackrel{}{\leq} \sfrac{3\ve}{2}. \label{eqn:WZ7}
\end{align}
Next, from \eqref{eqn:WZ2}, \eqref{eqn:WZ5} and Lemma~\ref{lem:UsefulLem1} of Section~\ref{Sec-ReqRes}, we see that:
\begin{align}
\Pr[ \mathring{K}' \neq g_{\mathcal C_0} (Y^n, \mathring K) ] &\leq \sfrac{3\ve}{2},\\
\Pr\big[W^n(\mathring K, \mathring K') \neq W^n(\mathring K,  g_{\mathcal C_0} (Y^n, \mathring K))  \big] &\leq \sfrac{3\ve}{2}. \label{eqn:WZ8}
\end{align}
Combining \eqref{eqn:WZ7} and \eqref{eqn:WZ8}, we conclude that 
\begin{align}
\hspace{-2mm}\Pr\big[( W^n(\mathring K,g_{\mathcal C_0} \hspace{-0.5mm}(Y^n\hspace{-0.5mm}, \mathring K)),X^n,Y^n) \hspace{-0.5mm}\notin\hspace{-0.5mm} T^n_\ve [ Q_{W|X}\mathsf Q_{XY}] \big]  \leq 3\ve. \label{eqn:WZ9}
\end{align}
Thus, if the receiver estimates $\mathring K'$ using $g_{\mathcal{C}_0}(Y^n, \mathring{K})$, and sets $\hat X_i \triangleq f(W_i(\mathring K, g_{\mathcal{C}_0}(Y^n,\mathring{K})), Y_i)$ as the reconstruction for $X_i$, $i=1,\ldots, n$, then with a probability of $1-3\ve$, the per-symbol distortion is at most $\Delta(1+3\ve)$, since
\begin{align}
\hspace{-2mm}\Pr\hspace{-0.5mm}\left[ d(X^n\hspace{-0.5mm},f(W^n(\mathring K, \mathring g_{\mathcal C_0} \hspace{-0.5mm}(Y^n\hspace{-0.5mm}, \mathring K)), Y^n)) \hspace{-0.75mm}>\hspace{-0.75mm} \Delta(1\hspace{-0.5mm}+\hspace{-0.5mm}\ve)\right] \hspace{-0.75mm}\leq\hspace{-0.5mm} 3\ve.\label{eqn:WZ10}
\end{align}
Thus, we are guaranteed to have an average per-symbol distortion of no more than $\Delta + 3\ve d_{\max}$. We are nearly done, if we ensure that:
\begin{itemize}
\item[(1)] a suitable  $Q_{K,K'\tilde X^n}^\textsf{\tiny{approx}}$ is selected; and 
\item[(2)] the encoding is deterministic. (The encoding above involves randomly generating $(\mathring K, \mathring K')$ using $Q_{K,K'\tilde X^n}^\textsf{\tiny{approx}}$.)
\end{itemize}
 We can guarantee the first requirement by invoking Lemma~\ref{lem-ListDec} of Section~\ref{Sec-ReqRes}, which ensures that an approximation $Q^\textsf{\tiny{approx}}_{ K,K', \tilde X^n}$ of $Q_{ K,K',\tilde X^n}$ meeting \eqref{eqn:ApproxQ} can be realized if the encoder is given a uniform random seed of rate $R+R'-I(X;W)+\ve=2\ve$ that is independent of $\tilde X^n$. We can ensure the second requirement by approximating this uniform seed by a near-uniform seed of rate $2\ve$ obtained as a function of $\{X_{n+\ell}: \ell =1,\ldots, \frac{3\ve n}{H(X)}\}$ that extracts its intrinsic randomness (Lemma~\ref{lem-IntrinsicRand} of Section~\ref{Sec-ReqRes}). Thus, both the pmf $Q_{ \mathring K,\mathring K', X^n}$ of \eqref{eqn:WZ4} and the encoding operation can be realized as a deterministic function of $n+\frac{3\ve n}{H(X)}$ symbols of the $X$ source. 

Finally, since the last $\frac{3\ve n}{H(X)}$ source symbols are used solely to generate the random seed, it can be  assumed that the average distortion corresponding to each of these symbols is no more than $d_{\max}$. Combining this with the estimate for the first $n$ symbols, we see that the overall average per-symbol distortion offered by the code is at most $ \Delta+3\ve d_{\max} + \frac{3\ve }{H(X)} d_{\max}$. The result then follows by limiting $\ve$ to zero.
\end{IEEEproof}

\subsection{Near-uniform Distributed Lossy Source Coding Problem}

We begin by analyzing joint near-uniformity of encoder outputs in a variant of the Slepian-Wolf (SW) problem, which will be used for the corresponding distributed lossy compression problem. Since the lossless compression of a source with near-uniform output is not possible without  shared randomness between encoder and decoder~\cite{Vellambi-Kliewer-Bloch-ISIT15}, SW coding with jointly near-uniform encoder outputs is also not possible. However, if we relax the decoder constraint to lossless recovery of all but a small fraction of symbols, then there exist distributed coding schemes with  jointly near-uniform encoder outputs provided the two sources share non-trivial G\'{a}cs-K\"{o}rner common information. The following result quantifies this precisely.
\begin{theorem}\label{thm:SW}
Let $(X,Y)\sim \mathsf Q_{XY}$ and suppose that the random variable $U$ common to $X$ and $Y$  (in the G\'{a}cs-K\"{o}rner sense) be non-trivial. Let $\ve\in(0,  H(U))$. Then, for any $(R_x,R_y)$ in the interior of the Slepian-Wolf rate region, there exist $n\in\mathbb N$ and $m\in \llbracket n, n+\frac{9\ve n}{H(U)} \rrbracket$, encoding functions $f_X:\X^{m} \rightarrow \llbracket 1, 2^{nR_x} \rrbracket $ and  $f_Y:\Y^{m} \rightarrow \llbracket 1, 2^{nR_y}\rrbracket $ operating over $m$ source symbols, and a decoding function $g_{XY}: \llbracket 1, 2^{nR_x}\rrbracket \times \llbracket 1, 2^{nR_y}\rrbracket \rightarrow \X^n\times \Y^n$ outputting $n$ symbols of both sources   such that
\begin{align}
\mathbb{V}(Q_{f_X(X^{m}), f_Y(Y^{m})}, \mathsf{unif} ( \llbracket 1, 2^{nR_x}\rrbracket \times \llbracket 1, 2^{nR_y}\rrbracket) ) &\leq \ve,\\
\Pr[(X^n,Y^n)\neq g_{XY}(f_X(X^{m}), f_Y(Y^{m}))] &\leq \ve.
\end{align}
\end{theorem}

\begin{IEEEproof}
The proof proves that the claim holds for a corner point of the SW rate region, which extends to the other corner point by reversing the roles of the sources, and to the interior of the rate region by time-sharing. Without loss of generality, let us build a coding scheme for the corner point at which $Y$ is available at the decoder. Let $U$ indicate the G\'{a}cs-K\"{o}rner common randomness between $X$ and $Y$. Let
\begin{align}
[R_u\,\, R_x \,\, R_y] &\triangleq [H(U)+\sfrac{\ve}{2}\,\,\,\, H(X|Y)+\sfrac{\ve}{2}\,\,\,\, H(Y|U)+\sfrac{\ve}{2}]\\
R_x' &\triangleq I(X;Y|U) + \sfrac{\ve}{2}< I(X;Y)\label{eqn-Rx'choice}
\end{align}
Thus, $R_u+R_y>H(U,Y) = H(Y)$, and $R_x>  H(X|Y)$.
  \begin{figure}[h!]
   \centering
   \includegraphics[scale=0.75]{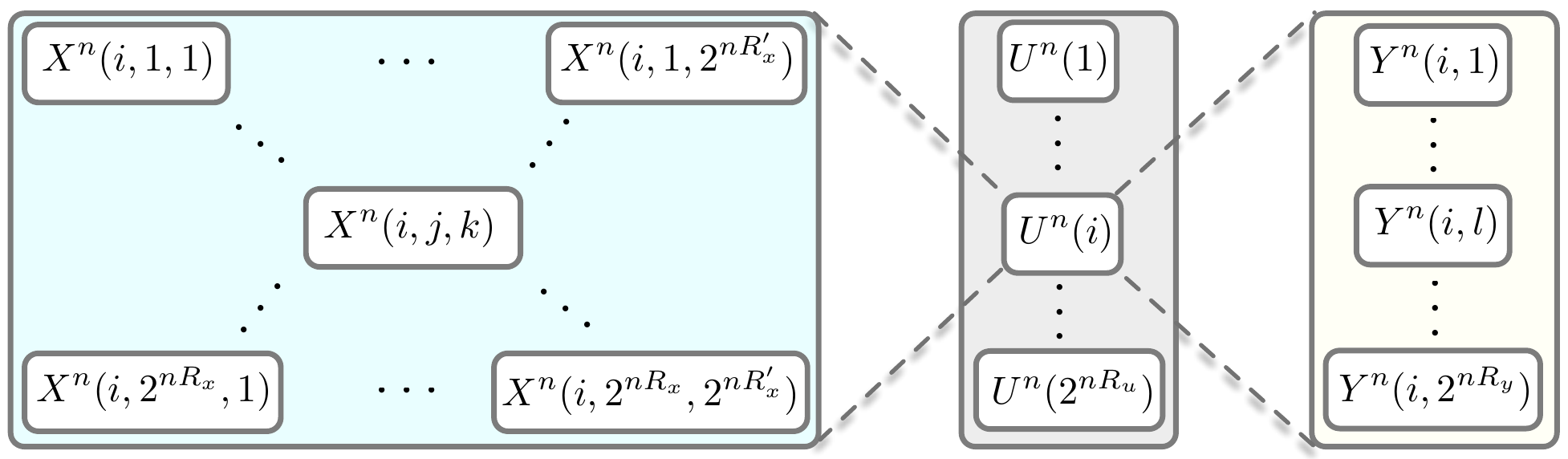}
   \caption{Codebook setup for the Slepian-Wolf problem}\label{Fig:3-SW}
 \end{figure}
 
 As illustrated in Fig.~\ref{Fig:3-SW}, a random codebook $\{U^n(1),\ldots, U^n(2^{nR_u})\}$ by choosing codewords  i.i.d. according to $\mathsf Q_U^{\otimes n}$. For each $i\in \llbracket 1, 2^{nR_U}\rrbracket$, generate a codebook of $X$-codewords arranged as $2^{nR_x}$ rows and $2^{nR_x'}$ columns with codewords selected i.i.d. using $\mathsf Q_{X|U}^{\otimes n}(\cdot |U^n(i))$. Note that this codebook has $2^{n(H(X|U)+\ve)}$ entries. Next, for each $i\in \llbracket 1, 2^{nR_U}\rrbracket$, generate a codebook of $2^{nR_y}$ $Y$-codewords with codewords selected i.i.d. using $\mathsf Q_{Y|U}^{\otimes n}(\cdot |U^n(i))$. We let $\mathcal{C}$ to jointly represent the three codebooks. Now, let random indices  $I, J,K,L$ satisfy
  \begin{align}
  Q_{I,J,K} &\triangleq  \mathsf{unif} (  \llbracket 1, 2^{nR_u}\rrbracket \times  \llbracket 1, 2^{nR_x}\rrbracket \times \llbracket 1, 2^{nR_x'}\rrbracket),\\
   Q_{I,L} &\triangleq \mathsf{unif} (  \llbracket 1, 2^{nR_u}\rrbracket \times  \llbracket 1, 2^{nR_y}\rrbracket).
  \end{align}
Let $ \hat{U}^n \triangleq  U^n(I)$, $\hat{X}^n \triangleq X^n (I,J,K)$ and $\tilde{Y}^n \triangleq Y^n(I,L)$,  and let $\hat{Y}^n $ be the output of the DMC $\mathsf Q_{Y|X}$ when the input is $ \hat{X}^n = X^n (I,J,K)$. By an  application of Lemma~\ref{lem-codeconst} of Section~\ref{Sec-ReqRes}, we see that 
 \begin{align}
\Exp_\mathcal{C} [D_{KL}(Q_{\hat{X}^n}\parallel \mathsf Q^{\otimes n}_{X})] &\mathop{\longrightarrow}^{n\rightarrow\infty} 0,\label{eqn-MTSCConseq2}\\
\Exp_\mathcal{C} [D_{KL}(Q_{\tilde{Y}^n}\parallel \mathsf  Q^{\otimes n}_Y)] &\mathop{\longrightarrow}^{n\rightarrow\infty}0,\label{eqn-MTSCConseq3}
\end{align}
since $R_{u}+R_x+R_x'>H(U,X)=H(X)$, $R_u+R_y>H(Y,U)=H(Y)$, and $R_u>H(U)$. Note that since $\hat{Y}^n$ and $Y^n$ are obtained by transmitting $\hat X^n$ and $X^n$, respectively, on the DMS $\mathsf Q_{Y|X}$, we are also guaranteed that 
 \begin{align}
\Exp_\mathcal{C} [D_{KL}(Q_{\hat{X}^n\hat Y^n}\parallel \mathsf Q^{\otimes n}_{XY})] &\mathop{\longrightarrow}^{n\rightarrow\infty} 0,\label{eqn-MTSCConseq2a}
\end{align}

Further, using a similar argument, we can also show that
  \begin{align}
\Exp_\mathcal{C} \left[2^{-nR_u}{ \sum_i} D_{KL}\big(Q_{\hat{Y}^n|I=i}\parallel \mathsf Q_{Y|U}^{\otimes n}(\cdot |U^n(i))\big)\right] &\mathop{\longrightarrow}^{n\rightarrow\infty}0\label{eqn-MTSCConseq4},\\
\Exp_\mathcal{C}\left[2^{-nR_u}{ \sum_i}D_{KL}\big(Q_{\tilde{Y}^n|I=i}\parallel\mathsf Q_{Y|U}^{\otimes n}(\cdot |U^n(i))\big)\right] &\mathop{\longrightarrow}^{n\rightarrow\infty}0\label{eqn-MTSCConseq4a}.
  \end{align}
Now, let $\eta_n \triangleq \Exp[D_{KL}( Q_{\hat{Y}^n, I,J} \parallel Q_{\hat Y ^n, I} Q_J)]$. Then, the following argument holds.
 \begin{align}
  \eta_n&= \Exp_{\mc C}\big[D_{KL}( Q_{\hat{Y}^n, I,J} \parallel Q_{\hat Y ^n, I} Q_J)\big]  \label{eqn-Analys1}\\
  &= \Exp_{\mc C}\hspace{-0.75mm}\Big[D_{KL}( Q_{\hat{Y}^n,I,J} \hspace{-0.75mm}\parallel\hspace{-0.75mm} Q_{I} Q_{J}\mathsf Q_{Y|U}^{\otimes n}(\cdot|U^n(I)))\hspace{-0.25mm}-\hspace{-0.25mm}D_{KL}( Q_{\hat{Y}^n,I} \hspace{-0.75mm}\parallel\hspace{-0.75mm} Q_{I} \mathsf Q_{Y|U}^{\otimes n}(\cdot|U^n(I))\Big]\\
   & \stackrel{(a)}{\leq} \Exp_{\mc C}\hspace{-1mm}\Big[D_{KL}( Q_{\hat{Y}^n,I,J} \hspace{-1mm}\parallel\hspace{-1mm} Q_{I,J}\mathsf Q_{Y|U}^{\otimes n}(\cdot|U^n(I)))\Big]\\
  & \stackrel{(b)}{=} \Exp_{\mc C}\hspace{-1mm}\left[\sum_{y^n}\bigg[ {\textstyle \frac{\sum_{k}\mathsf Q^{\otimes n}_{Y|X}(y^n|X^n(1,1,k))}{2^{nR_x'}}}\bigg]\log_2 \frac{\sum_{k'} {\mathsf Q^{\otimes n}_{Y|X}(y^n|X^n(1,1,k'))}}{{2^{nR_x'}} \mathsf Q_{Y|U}^{\otimes n}(y^n|U^n(1))}\right]\\
  &=\sum_{y^n,k}\Exp_{U^n(1),X^n(1,1,k)}\left[ \frac{\mathsf Q^{\otimes n}_{Y|X}(y^n|X^n(1,1,k))}{2^{nR_x'}}\Exp_{\textrm{rest}}\left[\log_2 \frac{\sum_{k'} {\mathsf Q^{\otimes n}_{Y|X}(y^n|X^n(1,1,k'))}}{ {2^{nR_x'}}\mathsf Q_{Y|U}^{\otimes n}(y^n|U^n(1))}\Bigg| \substack{U^n(1)\\X^n(1,1,k) }\right]\right] \label{eqn-Analys4}\\
   &\leq\sum_{y^n,k}\Exp_{U^n(1),X^n(1,1,k)}\left[ \frac{\mathsf Q^{\otimes n}_{Y|X}(y^n|X^n(1,1,k))}{2^{nR_x'}}\log_2\Exp_{\textrm{rest}}\left[ \frac{\sum_{k'} {\mathsf Q^{\otimes n}_{Y|X}(y^n|X^n(1,1,k'))}}{{2^{nR_x'}}\mathsf Q_{Y|U}^{\otimes n}(y^n|U^n(1))}\Bigg| \substack{U^n(1)\\X^n(1,1,k)}\right]\right]\label{eqn-Analys5}\\
    &\leq \sum_{y^n,k}\Exp_{U^n(1),X^n(1,1,k)}\left[ \frac{\mathsf Q^{\otimes n}_{Y|X}(y^n|X^n(1,1,k))}{2^{nR_x'}}\log_2\left[1+\frac{{\mathsf Q^{\otimes n}_{Y|X}(y^n|X^n(1,1,k))}}{2^{nR_x'}\mathsf Q_{Y|U}^{\otimes n}(y^n|U^n(1))}\right]\right] \label{eqn-Analys6}\\
    &\leq \log_2 \big( 1+ 2^{n(I(X;Y|U)-R_x'+2\delta \log_2 |\Y|)}\big) + 2|\X||\Y||\U|e^{-n\delta^2 \mu} \log_2 \left(1+ \mu^{-n}\right) \mathop{\longrightarrow}^{n\rightarrow\infty}0 \quad (\textrm{due to \eqref{eqn-Rx'choice}}), \label{eqn-Analys7}
  \end{align}
  where
  \begin{itemize}
\item $(a)$ follows by dropping the second non-negative term that is subtracted;
\item $(b)$ due to the i.i.d. construction of the random codebooks;
\item \eqref{eqn-Analys4} uses the law of iterated expectations, where $\Exp_{\textrm{rest}}$ is the expectation over all codewords except $(U^n(1),X^n(1,1,k))$;
\item  \eqref{eqn-Analys5} uses Jensen's inequality for the $\log$ function;
\item  \eqref{eqn-Analys6} because $X^n(1,1,k')\sim \mathsf Q_{X|U}^{\otimes n}(\cdot|U^n(1))$ for $k' \neq k$, and
\begin{align}
\Exp_{\textrm{rest}} \big[ \mathsf Q^{\otimes n}_{Y|X}(\cdot|X^n(1,1,k')) \big| X^n(1,1,k)\big] \hspace{-1mm}=\hspace{-1mm} \mathsf Q^{\otimes n}_{Y|U}(\cdot|U^n(1)),
\end{align}
since $X^n(1,1,k')$ is chosen using ${\prod\limits_{\ell=1}^n}\mathsf Q_{X|U}(\cdot|U_\ell(1))$; and
\item  finally, \eqref{eqn-Analys7} follows by splitting the outer sum depending on whether the realization of the codeword $X^n(1,1,k)$ and $y^n$ are jointly $\delta$-strongly letter typical, where $\delta < \sfrac{\ve}{(4\log_2 |\Y|)}$, and  
\begin{align}
\mu \triangleq \min\{\mathsf Q_{X,Y,U}(x,y,u): {(x,y,u)\in\supp(\mathsf Q_{X,Y,U})}\}.
\end{align}
\end{itemize}
Note that because of the choice of $R_x'$ in \eqref{eqn-Rx'choice}, the bound in \eqref{eqn-Analys7} approaches $0$ as $n\rightarrow \infty$.  From \eqref{eqn-MTSCConseq2}-\eqref{eqn-MTSCConseq4a} and \eqref{eqn-Analys7}, we conclude that there must exist for a sufficiently large $n$, a codebook $\mc C^*$ such that
   \begin{align}
\mathbb{V}(Q_{\hat{U}^n}, \mathsf Q^{\otimes n}_U) &<\ve\label{eqn-V1}\\
\mathbb{V}(Q_{\hat{X}^n\hat{Y}^n}, \mathsf Q^{\otimes n}_{XY}) &<\ve,\label{eqn-V2}\\
\mathbb{V}(Q_{\tilde{Y}^n}, \mathsf  Q^{\otimes n}_Y) &<\ve,\label{eqn-V3}\\
\mathbb{V}( Q_{\hat{Y}^n, I,J},Q_{\hat Y ^n, I} Q_J)&<\ve\label{eqn-V4},\\
\mathbb{V}( Q_{\hat{Y}^n, I,},Q_{\tilde Y ^n, I} )&<\ve\label{eqn-V4a}.
  \end{align}
  The code $\mc C^*$ induces two joint pmfs $Q^*_{I,J,K, \hat U^n, \hat X^n,  \hat Y^n}$ and $Q^*_{I,L,\hat{U}^n\tilde{Y}^n}$ for which $Q^*_{I,J,K}=Q_{I,J,K}$ and $Q^*_{I,L} = Q_{I,L}$. 
Since \eqref{eqn-V4a} holds, there must exist a joint pmf $Q^\dagger_{\hat{Y}^n,\tilde{Y}^n, I}$ over $\Y^n\times \Y^n \times \llbracket 1, 2^{nR_u}\rrbracket$ that optimally couples $(\hat Y^n, I)$ and $(\tilde Y^n, I)$ so that 
  \begin{align}
  \Pr[(\hat Y^n, I) \neq (\tilde Y^n, I)] \leq 2\ve. \label{eqn-hattildebridge}\end{align} 
  Further, since $\hat Y^n$ and $\tilde Y^n$ are generated from the same $U$-codebook, they share $\hat U^n$ as common randomness in the G\'{a}cs-K\"{o}rner sense. 
  Now, let $Q^\circ_{I,J,K,\hat X^n, \tilde Y^n, L}$ be the marginal pmf of $(I,J,K,\hat X^n, \tilde Y^n, L)$ obtained from
  \begin{align}
 Q^\circ_{I,J,K,\hat X^n, \hat Y^n, \tilde Y^n, L} \triangleq Q^*_{I,J,K, \hat X^n, \hat Y^n} Q^\dagger_{\tilde{Y}^n | \hat Y^n, I} Q^*_{L | I, \tilde {Y}^n}.\label{eqn-Qcircdefn}
  \end{align} 
    For the pmf $Q^\circ_{I,J,K,\hat X^n, \tilde Y^n, L}$, we can show the following:
  \begin{align}
 \mathbb{V}(\mathsf Q^{\otimes n}_{XY}, Q^\circ_{\hat{X}^n \tilde{Y}^n}) &\stackrel{ \eqref{eqn-V2}, \eqref{eqn-hattildebridge},\eqref{eqn-Qcircdefn}}{\leq} 3\ve,   \label{eqn-sourcevardist}\\ 
 \mathbb{V}( Q^\circ_{\tilde{Y}^n, I,J},Q^\circ_{\tilde Y ^n, I} Q_J)&\stackrel{ \eqref{eqn-V4}, \eqref{eqn-hattildebridge},\eqref{eqn-Qcircdefn}}{\leq}5\ve\label{eqn-newV4}, \\
 \mathbb{V}( Q^\circ_{\tilde{Y}^n, I,L},Q^*_{\tilde Y ^n, I,L} )&\,\,\,\quad\stackrel{  \eqref{eqn-hattildebridge}}{\leq}\quad\,\,2\ve \label{eqn-newV5},\,\, 
  \end{align} 
  Note that even though $Q^\circ_{\tilde{Y}^n, I,L}$ and $Q^*_{\tilde{Y}^n, I,L}$ could be different, we can still view $\tilde{Y}^n$ as being generated using the two-stage codebook by first choosing the $U$-codeword uniformly at random, and then the $Y$-codeword by selecting the index $L$ according to $Q^\circ_{L|I}$, which is only nearly uniform.  Lastly, since in $Q^\circ$, we have $L \mkv (I, \tilde Y^n) \mkv (J,K)$, we also have 
  \begin{align}
   \mathbb{V}( Q^\circ_{L, I,J},Q^\circ_{L, I} Q_J)&\stackrel{\eqref{eqn-newV4}, \eqref{eqn-Qcircdefn}}{\leq}5\ve\label{eqn-newV6}, \,\, \\  \mathbb{V}( Q^\circ_{L, I,J},Q^*_{L, I} Q_J)&\stackrel{\eqref{eqn-newV6}, \eqref{eqn-newV5}}{\leq}7\ve\label{eqn-newV7}.
    \end{align}
    Thus, under the law $Q^\circ$, $(I,L)$ and $J$ are jointly nearly-uniform. We now use an approach similar to the Wyner-Ziv case to build a code for the  problem at hand.
     \begin{itemize}
\item The $X$-encoder first generates  $I^\circ \sim Q^\circ_{I\mid\hat U^n}(\cdot | U^n)$, and then $(J^\circ\hspace{-1mm}, K^\circ)\hspace{-0.75mm}\sim\hspace{-0.75mm} Q^\circ_{J,K| \hat X^n\hspace{-0.5mm}, I}(\cdot | X^n\hspace{-1mm},  I^\circ)$. It  sends $J^\circ$ to the receiver;
\item The $Y$-encoder generates  $I^\circ \sim Q^\circ_{I|\hat U^n}(\cdot | U^n)$ that matches the index generated by the $X$-encoder, and then generates ${L}^\circ\sim Q^\circ_{L| \tilde Y^n, I}(\cdot | Y^n,  I^\circ)$. It sends $( I^\circ, L^\circ)$ to the receiver;
\item The decoder declares $Y^n(I^\circ, L^\circ)$ as the realization of $Y^n$. It then looks for an index $K$ such that $X^n(I^\circ, J^\circ, K)$ is jointly typical with $Y^n(I^\circ, L^\circ)$. With high probability, the search will yield a unique $K$ that matches $K ^\circ$, since $K\in\llbracket 1, 2^{nR_x'} \rrbracket$ and $R_x' < I(X;Y)$ (see \eqref{eqn-Rx'choice}).
\end{itemize}
  The above encoding and decoding operations emulate the following joint pmf of sources and indices:
 \begin{align}
 \mathsf Q^{\otimes n}_{XY} Q^\circ_{I|\hat U^n}(\cdot | U^n)Q^\circ_{J,K|\hat X^n, I}(\cdot | X^n, \cdot) Q^\circ_{L | \tilde Y^n, I}(\cdot | Y^n, \cdot).\label{eqn-emultedpmf}
 \end{align}
According to \eqref{eqn-sourcevardist}, the variational distance between the emulated pmf and $Q^\circ_{I,J,K,\hat X, \tilde Y, L}$ is no more than $3\ve$, which when combined with \eqref{eqn-newV7} implies that the variational distance of the emulated joint pmf of $(I^\circ, J^\circ, L^\circ)$ is at most $10\ve$ away from the jointly uniform pmf $Q_{L, I} Q_J$.  Lastly, as in the Wyner-Ziv case, we are  done if we approximate the randomized encoders by functions, for which we use near-uniform seeds derived from additional source symbols in the following manner.
\begin{itemize}
\item  At both encoders, we use $U_{n+1},\ldots, U_{n+\frac{3\ve n}{H(U)}}$ to obtain the same near-uniform random seed over $\llbracket 1, 2^{2n\ve}\rrbracket$ , and then use the  seed to approximate the random index selection according to $Q^\circ_{I\mid\hat U^n}(\cdot | U^n)$.
\item We use $X_{n+\frac{3\ve n}{H(U)}+1},\ldots, X_{n+\frac{6\ve n}{H(U)}}$ to obtain a near-uniform random seed over $\llbracket 1, 2^{2n\ve}\rrbracket$, and then use the seed to realize index selection according to $Q^\circ_{J,K|\hat X^n, I}(\cdot | X^n, \cdot)$.
\item We use $Y_{n+\frac{6\ve n}{H(U)}+1},\ldots, X_{n+\frac{9\ve n}{H(U)}}$ to obtain a near-uniform random seed over $\llbracket 1, 2^{2n\ve}\rrbracket$, and then use the seed to realize index selection according to $Q^\circ_{L | \tilde Y^n, I}(\cdot | Y^n, \cdot)$.
\end{itemize}
In the above, extracting random seeds and realizing index selections as a function of the random seed and the sources are done by invoking Lemmas~\ref{lem-ListDec}  and \ref{lem-IntrinsicRand} of Section~\ref{Sec-ReqRes}. 

Thus, for sufficiently large $n$, there exist codes that encode $n+\frac{9\ve n}{H(U)}$ source symbols into a  jointly nearly uniformly distributed pair of indices, using which the first $n$ symbols can be losslessly retrieved with high probability.
  \end{IEEEproof}
  
We are now ready to present our result pertaining to uniform lossy compression in the two-source distributed lossy source coding problem. Note that the proof does not require a  characterization of the underlying rate region.
\begin{theorem}
Given jointly correlated sources $(X,Y)\sim \mathsf Q_{XY}$ with non-trivial G\'{a}cs-K\"{o}rner common information, distributed lossy compression with jointly near-uniform encoder outputs is possible at all rate points in the strict interior of the distributed lossy compression rate region.
\end{theorem}
\begin{IEEEproof}
Let $(R_x, R_y)$ be in the interior of the distributed lossy compression rate region. Fix $\ve>0$. Then, for sufficiently large $n$, there exist encoders $f_X$ and $f_Y$ operating at rates no more than $R_x+\ve$ and $R_y+\ve$, and a reconstruction function $g_{XY}$ that operates on the encoder outputs to generate reconstructions for $X$ and $Y$ with an average per-symbol distortion of at most $\Delta_x+\ve$ and $\Delta_y+\ve$, respectively. Without loss of generality, we may assume that $f_X(X^n)$ and $f_Y(Y^n)$ share non-trivial  G\'{a}cs-K\"{o}rner common information. Else, we can increase the encoded message rates by $\ve$ by appending to each encoder output, a function of $U$ -- the random variable common to $X$ and $Y$ in the G\'{a}cs-K\"{o}rner sense. 

Now, let $\mathring{X}=f_X(X^n)$ and $\mathring Y = f_Y(Y^n)$, and let $\mathring U$ be the random variable common to $\mathring X$ and $\mathring Y$ in the G\'{a}cs-K\"{o}rner sense. From Theorem~\ref{thm:SW}, we see that there exists sufficiently large $N\in\mathbb{N}$, sufficiently small $\delta$, and $M\leq N+ 9\delta N$ such that there exists a code that encodes $M$ symbols of the correlated source $(\mathring X, \mathring Y)$ in any interior point of its SW rate region and recovers the first $N$ source symbols of $\mathring X$ and $\mathring Y$ losslessly. Concatenating $M$ copies of the lossy source code with encoders $f_X$ and $f_Y$ (as the outer code) followed by the above code for $\mathring X$ and $\mathring Y$ (as the inner code) will yield a joint code operating at rates of no more than $R_x+2\ve+\delta$ and $R_y+2\ve+\delta$, respectively. Moreover, the average distortions offered by this joint code for the $nM$ symbols of $X$ and $Y$ are at most $\frac{\Delta_x+\ve+9\delta {d_x}_{\max}}{1+9\delta}$ and $\frac{\Delta_y+\ve+9\delta {d_y}_{\max}}{1+9\delta}$, respectively. Since $\ve$ and $\delta$ are arbitrary, the claim holds.
\end{IEEEproof}

\section{Required Results}\label{Sec-ReqRes}
\begin{lemma}\label{lem:UsefulLem1}
Let p.m.f. $Q_{A,B}$ over a finite set $\A\times \B$ be such that for $(A,B)\sim Q_{AB}$, there exists a function $\phi(B)$ such that $\Pr[A\neq \phi(B)] \leq \ve$. Now, let $(\tilde{A},\tilde{B})\sim \tilde Q_{\tilde A, \tilde B}$ be such that $\mathbb{V}( \tilde Q_{\tilde A, \tilde B}, Q_{A,B}) \leq \ve$. Then, $\Pr[ \tilde A \neq \phi(\tilde B)] \leq 2\ve$. 
\end{lemma}
\begin{IEEEproof}
Let $\mc S=\{(a,b): a\neq \phi(b)\}$. Then,
\begin{align}
 \Pr[A\neq \phi(B)]  & = Q_{A,B}(\mc S)\\
 \Pr[\tilde A\neq \phi(\tilde B)] & = \tilde Q_{\tilde A, \tilde B}(\mc S).
\end{align}
Thus,
\begin{align}
|\Pr[\tilde A\neq \phi(\tilde B)] &\leq  \Pr[A\neq \phi(B)] | + | \tilde Q_{\tilde A, \tilde B}(\mc S) - Q_{A,B}(\mc S)| \notag \\
&\leq \ve+ \sum_{(a,b)\in \mc S} | \tilde Q_{\tilde A, \tilde B}(a,b) - Q_{A,B}(a,b)| \\
&\leq \ve + \mathbb{V}( \tilde Q_{\tilde A, \tilde B}, Q_{A,B}) \leq 2\ve.
\end{align}
\end{IEEEproof}
\begin{lemma}\label{lem-ListDec}
Given p.m.f. $Q_{AB}$ over a finite alphabet $\A\times\B$ and $R>I(A;B)$, suppose that we construct a random codebook $\mathcal{C}_n$ of $2^{nR}$ $A$-codewords generated randomly using $Q_A$. Let $L\sim \mathsf{unif}(\llbracket 1, 2^{nR} \rrbracket)$. Suppose that $A^n(L)$ is transmitted over the DMC $Q_{B|A}$ and $\tilde{B}^n$ is the corresponding output. Let $S\sim \mathsf{unif}(\llbracket 1, 2^{n\rho} \rrbracket)$, where $\rho>R-I(A;B)$. Then, there exists $\phi_{\mathcal{C}_n}: \B^n \times \llbracket 1, 2^{n\rho} \rrbracket \rightarrow \llbracket 1, 2^{nR} \rrbracket $ (that depends on $\mathcal C_n$) such that
\begin{align}
\lim_{n\rightarrow \infty} \Exp\big[\mathbb{V}(Q_{\phi(\tilde B^n, S),\tilde B^n}, Q_{L,\tilde B^n})\big] = 0,
\end{align}
where $Q_{L,\tilde B^n}$ is the joint p.m.f. of $(L,\tilde{B}^n)$ induced by $\mathcal C_n$. 
\end{lemma}
\begin{IEEEproof}
Let $\delta, \ve >0$ be chosen such that 
\begin{align}
\rho-R+ I(A; B) - 4\delta \log_2 (|\mc A||\B|) > \ve. \label{eqn-infitineschoice}
\end{align}
By the random codebook construction, it follows that $(A^n(L),\tilde B^n)$ is as if it is the output from a DMS $Q_{AB}$. Hence, by \cite[Theorem 1.1]{Kramer-MUIF}, it follows that
\begin{align}
\Pr\left[ (A^n(L), \tilde{B}^n) \notin T_{\delta}^n[Q_{AB}]\right] \leq 2M e^{-n \delta^2 \mu},\label{eqn-doubleensembleave1}
\end{align}
where $M\triangleq |\mc A||\B|$ and  $\mu \triangleq \min\limits_{(a,b)\in\supp(Q_{AB})} Q_{AB}(a,b).$
Now, let for a codebook $\mc C_\circ \triangleq \{(a^n(l)\}_{l\in \llbracket 1,2^{nR}\rrbracket}$ and $b^n\in\B^n$,
 \begin{align}
 \mathcal L_\mathcal{C_\circ}(b^n) \triangleq \big\{ l: (a^n(l),b^n)\in  T_{\delta}^n[Q_{AB}] \big\}.\label{eqn-listsize}
  \end{align}
  From Lemma \ref{lem-listdecbnd} below, the following holds for sufficiently large $n$.
  \begin{align}
\Exp\left[ | \mathcal L_{\mathcal{C}}(\tilde B^n)|\right] \leq 2^{1+n\left(R - I(A;B) +2\delta \log_2 M\right)}.
  \end{align}
  Let $\mc F$ be the collection of codebooks $\mc C_0 \triangleq \{a^n(l)\}_{l\in \llbracket 1,2^{nR}\rrbracket}$ such that the following hold.
 \begin{align}
\Pr\left[ (a^n(L),\tilde{B}^n) \notin T_{\delta}^n[Q_{AB}]\,\big|\,\mc C =\mc C_0\right] &\leq \sqrt{2M e^{-n \delta^2 \mu}}\\
\frac{\Exp\big[ | \mathcal L_{\mathcal{C}}(\tilde B^n)| \big| \mc C=\mc C_0\big]}{ \Exp\big[ | \mathcal L_{\mathcal{C}}(\tilde B^n)|\big]} &\leq2^{\delta \log_2 M}.  \label{eqn-singleensembleave1}
 \end{align}
By Markov's inequality, we then have
 \begin{align}
 \Pr[ \mc C \notin \mc F] \leq \sqrt{2M e^{-n \delta^2 \mu}}+2^{-n\delta \log_2 M}. \label{eqn-mcFbnd}
 \end{align}
 Now, pick $\mc C^* \triangleq \big\{{a^*}^n(l)\big\}_{l\in \llbracket 1,2^{nR}\rrbracket} \in \mc F$ and define $\mc G_{\mc C^*}$ as the set of all $ b^n$ such that 
 \begin{align}
 \Pr\left[ ({a^*}^n(L), \tilde B^n )\notin T_{\delta}^n[Q_{AB}] \Big|\, {\setstretch{0.5}\begin{array}{rl} \tilde{B}^n &\hspace{-3mm}= b^n\\ \mc C&\hspace{-3mm}= \mc C^*\end{array}}\right]  &\leq \sqrt[4]{2M e^{-n \delta^2 \mu}} \\  | \mathcal L_{\mathcal{C}^*}(b^n)| &\leq 2^{1+n\left(R - I(A;B) +4\delta \log_2 S\right)}. \label{eqn-GC*defn}
 \end{align}
 Again, by Markov's inequality, it follows that
 \begin{align*}
 \Pr[ \tilde{B}^n \notin  \mc G_{\mc C^*} \mid \mc C=\mc C^* ] \leq \eta_0\triangleq \sqrt[4]{2M e^{-n \delta^2 \mu}}+2^{-n\delta \log_2 M}.
 \end{align*}
 Further, it also follows that for each $b^n\in \mc G_{\mc C^*}$,
 \begin{align}
 \sum_{l\notin  \mathcal L_{\mathcal{C}^*}(b^n)} Q_{L \hat{B}^n} (l, b^n) \stackrel{\eqref{eqn-GC*defn}}{\leq} \sqrt[4]{2M e^{-n \delta^2 \mu}}.
 \end{align}
 Thus by Lemma \ref{lem-randpmfselec}, we see that given a random seed $S\sim \mathsf{unif} (\llbracket 1, 2^{n \rho} \rrbracket)$ for all $b^n\in \mc G_{\mc C^*}$, we can construct $f_{b^n}:\llbracket 1, 2^{n\rho} \rrbracket \rightarrow  \llbracket 1,2^{nR}\rrbracket $ with 
 \begin{align}
\parallel Q_{f_{b^n}(S)} - Q_{L |\tilde{B}^n=b^n} \parallel_1&\leq \frac{|\mc L_{C^*}(b^n)|}{2^{n\rho}} +  \sqrt[4]{2M e^{-n \delta^2 \mu}}\\
 &\hspace{-1mm}\stackrel{\eqref{eqn-GC*defn}}{\leq} 2^{1+n\left(R - I(A;B)+4\delta\log_2 M-\rho\right)}+  \sqrt[4]{2M e^{-n \delta^2 \mu}}\\
 &\hspace{-1mm}\stackrel{\eqref{eqn-infitineschoice}}{\leq} \eta\triangleq {2^{1-n\ve}+  \sqrt[4]{2Me^{-n \delta^2 \mu}}}.  \label{eqn:L1-bnd}
 \end{align}
We can now glue these functions to define 
 \begin{align}
 \Lambda_{\mc C^*} (b^n, S) \triangleq \left\{ \begin{array}{ll} f_{b^n}(S) & b^n \in  \mc G_{\mc C^*}\\ {l^*}^k &  b^n \notin  \mc G_{\mc C^*} \end{array}\right.,
 \end{align}
 where ${l^*} \in  \llbracket 1,2^{n\rho}\rrbracket $. 
 By construction, for the selected code $\mc{C}^*$, we now have 
 \begin{align}
  \sum\limits_{b^n\in \mc G_{\mc C^*}} Q_{\tilde{B}^n}( b^n ) \parallel Q_{ \Lambda_{\mc C^*} (b^n, S) } - Q_{L |\tilde{B}^n=b^n} \parallel_1 &\leq \eta,\notag\\
 \sum\limits_{b^n\in \B^n} Q_{\tilde{B}^n}( b^n ) \parallel Q_{ \Lambda_{\mc C^*} (b^n, S) } - Q_{L |\tilde{B}^n=b^n} \parallel_1  &\leq \eta+2\hspace{0.2mm}\eta_0. \notag
 \end{align}
 
Since the RHS does not depend on the choice of $\mc C^*$ in $\mc F$,  
 \begin{align*}
 \Exp \left[ \parallel Q_{ \Lambda_{\mc C} (\tilde B^n, S), \tilde B^n} - Q_{L, \tilde B^n} \parallel_1 \mid \mc C \in \mc F\right] &\leq  \eta+2\hspace{0.2mm}\eta_0.
 \end{align*}
Next. using the fact that the variational distance between two p.m.f.s is at most 2, we also have
 \begin{align*}
 \Exp\left[ \parallel Q_{ \Lambda_{\mc C} (\tilde B^n, S), \tilde B^n} - Q_{L, \tilde B^n} \parallel_1 \mid \mid \mc C \notin \mc F \right]\ \leq  2\hspace{0.2mm}\Pr[\mc C \notin \mc F ].
\end{align*}
Finally, combining the above two equations and using \eqref{eqn-mcFbnd} completes the claim.
\end{IEEEproof}

\begin{lemma}\label{lem-listdecbnd}
Consider the setup of Lemma~\ref{lem-ListDec}. Let $\mc{L}_{\mc C}(\cdot)$ be as defined in \eqref{eqn-listsize}. Then, for $n$ large, 
  \begin{equation}
\Exp\left[ | \mathcal L_{\mathcal{C}}(\tilde B^n)|\right] \leq  2^{1+n\left(R - I(A;B) -2\delta \log_2 \left(|\mc A||\B|\right)\right)}.
  \end{equation}
\end{lemma}
\begin{IEEEproof}
Owing to the random codebook construction, 
\begin{align}
\Exp\left[ | \mathcal L_{\mathcal{C}}(\tilde B^n)|\right] &= \Exp\left[ | \mathcal L_{\mathcal{C}}(\tilde B^n)| \,\big|\, L=1\right]\\
&= \sum_{l} \Exp \left[ \mathds{1}\{l \in \mc L_\mathcal{C}(\tilde B^n)\} \,\big|\, L=1\right].
\end{align}
Since the codewords are chosen randomly, it follows that $\Exp \left[ \mathds{1}\{l \in \mc L_\mathcal{C}(\tilde B^n)\} | L=1\right]$ is the same for $l>2$. Hence,
\begin{align}
\Exp\left[ | \mathcal L_{\mathcal{C}}(\tilde B^n)|\right] &\leq 1+ (2^{n R }-1)\Exp \left[ \mathds{1}\{2 \in \mc L_\mathcal{C}(\tilde B^n)\} | L=1\right]. \label{eqn-allterms}
\end{align}
Clearly, $\Exp \left[ \mathds{1}\{2 \in \mc L_\mathcal{C}(\tilde B^n)\} | L=1\right]$ is exactly the probability that  realizations  $A^n\sim Q_{A}^{\otimes n}$, $B^n\sim Q_{B}^{\otimes n}$ selected independent of one another are jointly $\delta$-letter typical. Thus, by \cite[Theorem 1.1]{Kramer-MUIF}, it follows that 
\begin{align*}
\Exp \left[ \mathds{1}\{2 \in \mc L_\mathcal{C}(\tilde B^n)\} | L=1\right] &=\sum_{(a,b^n)\in T_{\delta}^n[Q_{AB}]} Q_{A}(a^n) Q_B(b^n)  \leq 2^{-n(I(A;B) - 2\delta \log_2 |\mc A||\B|)}.
\end{align*}
Combining the above bound with \eqref{eqn-allterms} completes the proof.
\end{IEEEproof}
\begin{lemma} \label{lem-randpmfselec}
Let $Q$ be a p.m.f. on a finite set $\A$ such that there exists $\B\subseteq \A$ with $|\B|=M$ and $\sum_{b\in\B} Q(b) \geq 1-\ve$ for  $0<\ve<1$. Now, suppose that $L\sim \mathsf{unif} (\llbracket 1, \ell \rrbracket)$. Then, there exists $f: \llbracket 1, \ell \rrbracket \rightarrow \A$ such that $Q_{f(L)}$, the p.m.f. of $f(L)$, satisfies $\parallel Q_{f(L)} - Q \parallel_1 \leq \ve+ \frac{M}{\ell}$.
\end{lemma}
\begin{IEEEproof}
Let $b_1 \preceq b_2 \preceq \cdots \preceq b_M$ be an ordering of $\B$. Let $p_0 = 0$, and for $1\leq i\leq M$, let $p_i \triangleq \sum_{j=1}^i Q(b_j)$ denote the cumulative mass function. Now, let $N_i \triangleq \left\lfloor {p_i}\ell \right\rfloor$, $i=0,\ldots, M$, and let $f:\left\llbracket 1, N_M \right\rrbracket \rightarrow \B$ be defined by the pre-images via $f^{-1}(b_i) = \{ N_{i-1}+1,\ldots, N_i\}$, $i=1,\ldots, M$.
Fig.~\ref{Fig-8} provides an illustration of these operations. Now, by construction, we have
\begin{align}
0\leq p_i - \Pr\left[f(L) \in \{b_1,\ldots, b_i\} \right] \leq\ell^{-1}, \quad i=1,\ldots, M.
\end{align}

\begin{figure}[!h]
\centering
 \includegraphics[width=5in]{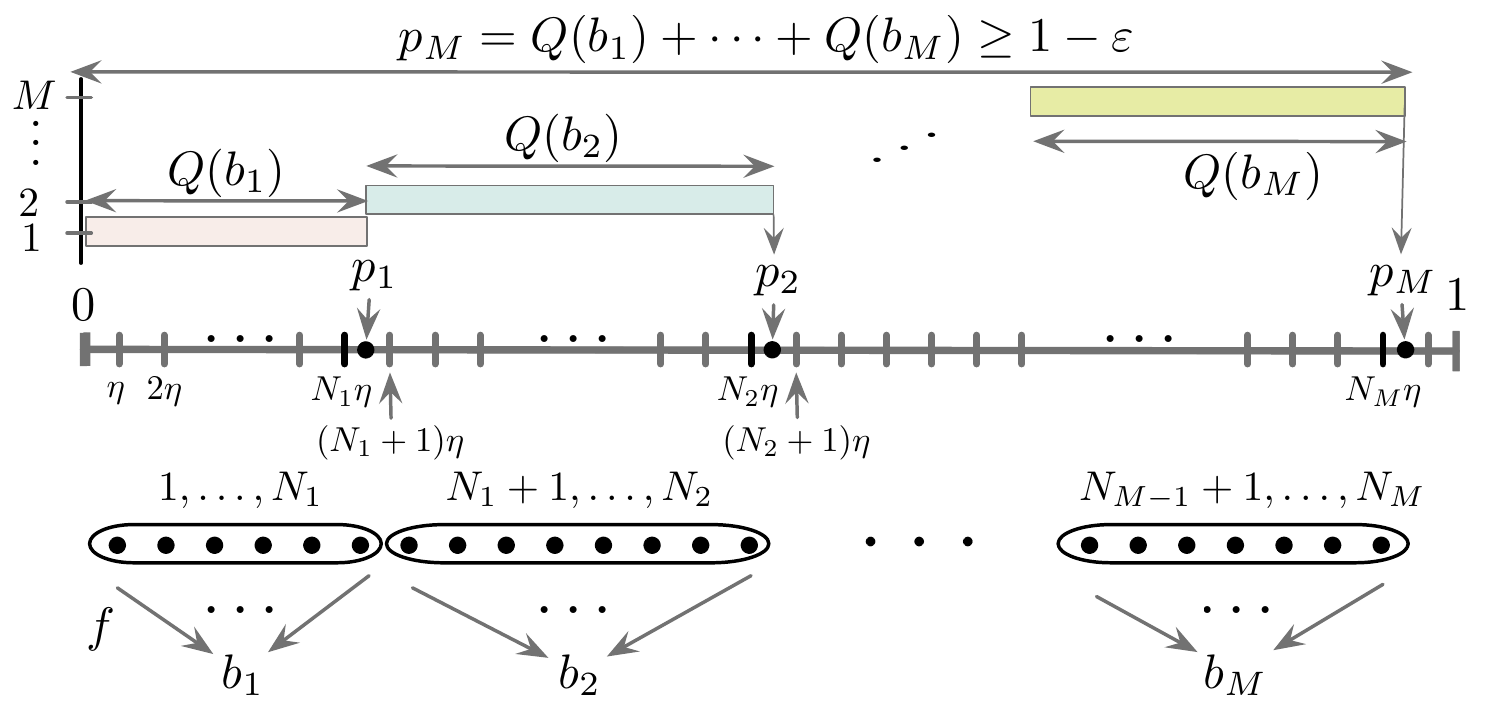}
\caption{An illustration of approximating a p.m.f. using a function of a uniform RV.}
 \label{Fig-8}
\end{figure}
Consequently, we also have for any $i=1,\ldots, M$,
\begin{align}
-\ell^{-1} \leq p_{i} -p_{i-1} - Q_{f(L)}(b_{i}) = Q(b_{i}) -Q_{f(L)}(b_{i}) \leq \ell^{-1}. \label{eqn-pmfapproxbnd}
\end{align}
Hence, we see that
\begin{align}
\sum_{a\in \A} | Q(a)- Q_{f(L)}(a) | &= \sum_{i=1}^M | Q(b_i)- Q_{f(L)}(b_i) |  + \Pr[ A\notin \B]\stackrel{\eqref{eqn-pmfapproxbnd}}{\leq}  \frac{M}{\ell} +\ve.
\end{align}
\end{IEEEproof}
\begin{lemma}\label{lem-codeconst}
Given pmf $p_{AB}$ and rates $R_A,R_B\in[0,\infty)$ such that $R_B> I(C;B)$ and $R_A+R_B>I(C;A,B)$, let us construct a random codebook $\{B^n(1),\ldots, B^n(2^{nR_B})\}$ with codewords chosen i.i.d. using $p_B^{\otimes n}$. For each $i\in\llbracket 1,2^{nR_B}\rrbracket$, generate a random codebook $\{A^n(i,1),\ldots, A^n(i,2^{nR_A})\}$ with codewords chosen i.i.d. using $ Q_{A|B}^{\otimes n}(\cdot | B^n(i))$. Let $(I,J)\sim\mathsf{unif}(\llbracket 1,2^{nR_B}\rrbracket\times \llbracket 1,2^{nR_A}\rrbracket)$, and let $\hat C^n$ be the output when $(A^n(I,J), B^n(I))$ is sent over the DMC $Q_{C|AB}$. Then,
\begin{align}
\Exp [D_{KL}(Q_{\hat{C}^n}\parallel \mathsf  Q^{\otimes n}_C)] &\mathop{\longrightarrow}^{n\rightarrow\infty}0,
\end{align}
where the expectation is over all the codebook realizations.
\end{lemma}
\begin{IEEEproof}
The proof follows from the achievability scheme and (10)-(15) in~\cite{Vellambi-Kliewer-Bloch-ITW15} by setting $\h=2$, $A_0=B$, $A_1=A$, $X_1=C$, and $B_1=B_2=X_2=\textrm{const}$.
\end{IEEEproof}

\begin{lemma}[Theorem 2.2.2 \cite{TSH-Book}] 
Let $X^n$ be i.i.d. according to $p_X$. Then, for each $R<H(X)$, there exists a sequence of mappings  $\{\phi_{n,R}:\X^n \rightarrow\llbracket 1, 2^{nR} \rrbracket\}_{n\in\mathbb{N}}$ such that 
\begin{align}
\lim_{n\rightarrow\infty} \mathbb{V}\big(\phi_{n,R}(X^n), \mathsf{unif}(\llbracket 1, 2^{nR} \rrbracket)\big) = 0.
\end{align} \label{lem-IntrinsicRand}
\end{lemma}


\begin{thebibliography}{1}
\providecommand{\url}[1]{#1}
\csname url@samestyle\endcsname
\providecommand{\newblock}{\relax}
\providecommand{\bibinfo}[2]{#2}
\providecommand{\BIBentrySTDinterwordspacing}{\spaceskip=0pt\relax}
\providecommand{\BIBentryALTinterwordstretchfactor}{4}
\providecommand{\BIBentryALTinterwordspacing}{\spaceskip=\fontdimen2\font plus
\BIBentryALTinterwordstretchfactor\fontdimen3\font minus
  \fontdimen4\font\relax}
\providecommand{\BIBforeignlanguage}[2]{{%
\expandafter\ifx\csname l@#1\endcsname\relax
\typeout{** WARNING: IEEEtran.bst: No hyphenation pattern has been}%
\typeout{** loaded for the language `#1'. Using the pattern for}%
\typeout{** the default language instead.}%
\else
\language=\csname l@#1\endcsname
\fi
#2}}
\providecommand{\BIBdecl}{\relax}
\BIBdecl

\bibitem{Han-IT2005}
T.~S. Han, ``Folklore in source coding: Information-spectrum approach,''
  \emph{IEEE Transactions on Information Theory}, vol.~51, no.~2, pp. 747--753,
  February 2005.

\bibitem{Hayashi-IT2008}
M.~Hayashi, ``Second-order asymptotics in fixed-length source coding and
  intrinsic randomness,'' \emph{IEEE Transactions on Information Theory},
  vol.~54, no.~10, pp. 4619--4637, October 2008.

\bibitem{Chou-Bloch-ISIT13}
R.~Chou and M.~Bloch, ``Data compression with nearly uniform output,''
  \emph{2013 IEEE International Symposium on Information Theory}, pp.
  1979--1983, 2013.

\bibitem{Vellambi-Kliewer-Bloch-ISIT15}
B.~N. Vellambi, M.~Bloch, R.~Chou, and J.~Kliewer, ``Lossless and lossy source
  compression with near-uniform output: Is common randomness always required?''
  \emph{2015 IEEE International Symposium on Information Theory}, pp.
  2171--2175, 2015.

\bibitem{Kontoyiannis-PtwiseRedundancy}
I.~Kontoyinannis, ``Pointwise redundancy in lossy data compression and
  universal lossy data compression,'' \emph{IEEE Transactions on Information
  Theory}, vol.~46, no.~1, pp. 136--152, January 2000.

\bibitem{TSH-Book}
T.~S. Han, \emph{Information-Spectrum Methods in Information Theory},
  1st~ed.\hskip 1em plus 0.5em minus 0.4em\relax Springer, 2003.

\bibitem{Cuff-LikelihoodEnc}
P.~Cuff and E.~Song, ``The likelihood encoder for source coding,'' in
  \emph{2013 IEEE Information Theory Workshop}, Sept 2013, pp. 1--2.

\bibitem{Kramer-MUIF}
G.~Kramer, ``Topics in multi-user information theory,'' \emph{Found. Trends
  Commun. Inf. Theory}, vol.~4, no. 4-5, pp. 265--444, 2007.
  
  \bibitem{Vellambi-Kliewer-Bloch-ITW15}
B.~N. Vellambi, J.~Kliewer, and M.~Bloch, ``Strong coordination over multihop
  line networks,'' \emph{2015 IEEE Information Theory Workshop}, pp. 192--196,
  2015.

\end{thebibliography}
\end{document}